\documentclass{aa}  

\usepackage{graphicx}
\usepackage{float}
\usepackage{txfonts}
\usepackage{orcidlink}
\usepackage{multirow}
\usepackage[utf8]{inputenc}
\usepackage{natbib}
\usepackage{booktabs}
\usepackage{longtable}
\usepackage[flushleft]{threeparttable}
\usepackage{tabularx}
\usepackage{makecell}
\usepackage[fleqn]{amsmath}
\usepackage{hyperref}
\usepackage{ulem}
\usepackage{wrapfig}
\usepackage{upgreek}
\hypersetup{
  colorlinks   = true,
  urlcolor     = blue,
  linkcolor    = blue,
  citecolor    = blue
}
\bibpunct{(}{)}{;}{a}{}{,} 
\newcommand{\psr}{NGC~6749A}
\newcommand{\msun}{\rm M_{\odot}}

\newcommand{\tempo}{\texttt{TEMPO}}
\newcommand{\Dracula}{\texttt{Dracula}}

\newcommand{\dmunit}{$\rm cm^{-3}\, pc$}
\newcommand{\mura}{$\mu_{\alpha}$}
\newcommand{\mudec}{$\mu_{\delta}$}
\newcommand{\muraGC}{$\mu_{\alpha, \rm GC}$}
\newcommand{\mudecGC}{$\mu_{\delta, \rm GC}$}
\newcommand{\masy}{\rm mas\,yr$^{-1}$}

\newcommand{\pbdot}{$\dot{P}_\mathrm{b}$}
\newcommand{\kms}{$\rm km \, s^{-1}$}

\defcitealias{Ridolfi_19}{R19}
\defcitealias{Joshi_Rasio_1997}{JR97}

\begin{document}

\title{The one and the only: the pulsar - white dwarf system in NGC 6749}
\titlerunning{The pulsar - white dwarf system in NGC 6749}

   \author{   
           Paulo~C.~C. Freire\orcidlink{0000-0003-1307-9435}\inst{1}
           Yinfeng Dai\orcidlink{0009-0007-6396-7891}\inst{2,3},
           Mario Cadelano\orcidlink{0000-0002-5038-3914}\inst{4,5},
           Cristina Pallanca\orcidlink{0000-0002-7104-2107}\inst{4,5},
           Zurong Zhou\orcidlink{0009-0005-8224-0677}\inst{6,7},
           Zhichen Pan\orcidlink{0000-0001-9754-9777}\inst{8,9,10,11},
           Luca Rosignoli\orcidlink{0000-0002-0327-5929}\inst{4},
           Davide Massari\orcidlink{0000-0001-8892-4301}\inst{5},
           Mattia Libralato\orcidlink{0000-0001-9673-7397}\inst{12},
           Craig Heinke\orcidlink{0000-0003-3944-6109}\inst{13}
          }

   \institute{Max-Planck-Institut f{\"u}r Radioastronomie, Auf dem H{\"u}gel 69, D-53121 Bonn, Germany
   \and
   School of Physics and Astronomy, Beijing Normal University, Beijing 100875, China
   \and
   Department of Physics, Faculty of Arts and Sciences, Beijing Normal University, Zhuhai 519087, China
   \and
   Dipartimento di Fisica e Astronomia "Augusto Righi", Alma Mater Studiorum Universit\`a di Bologna, via Piero Gobetti 93/2, I-40129 Bologna, Italy
   \and
   INAF - Osservatorio di Astrofisica e Scienze dello Spazio di Bologna, Via Piero Gobetti 93/3, I-40129 Bologna, Italy
   \and
   National Time Service Center, Chinese Academy of Sciences, Xi'an 710600, China
   \and
   Key Laboratory of Time Reference and Applications,Chinese Academy of Sciences, Xi'an 710600, China
   \and
   National Astronomical Observatories, Chinese Academy of Sciences, 20A Datun Road, Chaoyang District, Beijing 100101, People's Republic of China
   \and
   Guizhou Radio Astronomical Observatory, Guizhou University, Guiyang 550025, People’s Republic of China
   \and 
   Key Laboratory of Radio Astronomy and Technology, National Astronomical Observatories, Chinese Academy of Sciences, Beijing 100101, People's Republic of China
   \and
   College of Astronomy and Space Sciences, University of Chinese Academy of Sciences, Beijing 100049, People’s Republic of China
   \and
   Istituto Nazionale di Astrofisica, Osservatorio Astronomico di Padova, Vicolo dell'Osservatorio 5, Padova, IT-35122, Italy
   \and
   Department of Physics, University of Alberta, CCIS 4-183, Edmonton, AB T6G 2E1, Canada
   }
\authorrunning{Freire et al.}
   \date{Received --; accepted --}

\abstract{
PSR~J1905+0154A is a binary millisecond pulsar located in the globular cluster (GC) NGC~6749. It was discovered in 2004 in a search for pulsars in GCs carried out with the Arecibo 305-m radio telescope. The pulsar has a spin period of 3.2 ms, an orbital period of 0.81 days, and is in a low-eccentricity orbit with a low-mass WD companion.
Combining timing data from the early Arecibo observations of NGC 6749 with timing data from observations this GC made with the Five Hundred meter Aperture Spherical Telescope (FAST), we were able to derive, for the first time, a phase-coherent timing solution for this pulsar, which now spans 20 years. This includes a precise measurement of the astrometric, spin and orbital parameters of the system.
The small range of predicted accelerations expected from the gravitational field of this GC allows an estimate of the intrinsic spin-down: the inferred magnetic field at the surface (2.2 - 2.4 $\times 10^8\, $G) and characteristic age (2.8 - 3.5 Gyr) are typical of what one finds among MSPs in the Galactic field.
The position of this pulsar, the only confirmed to date in this GC, coincides with the position of one of the very few candidate white dwarfs (WDs) in the whole HST dataset on this GC. The position of the companion in the colour-magnitude diagram is consistent with a Helium WD with a mass of 0.17 - $0.19\, \msun$ (which implies an orbital inclination between 28 and 40 degrees), a cooling age of 0.4 - 0.7 Gyr, and a surface temperature of 11,600 - 14,800 K. A comparison with the characteristic age of the pulsar indicates that at the time of Roche lobe detachment, the spin period was between 1.98 and
2.62 ms. The relatively large proper motion difference relative to the motion of the GC, which is 4.5-$\sigma$ significant and an order of magnitude larger than the escape velocity, raises the possibility that, despite its location close to the centre of the GC, the pulsar might not be associated with it. Finally, our effort to confirm a second pulsar candidate in this GC did not yield a positive confirmation, nor the discovery of any additional pulsar in this GC.}

\keywords{Pulsars -- Globular clusters -- frequency derivatives}

\maketitle
   
\section{Introduction}

Globular clusters (GCs) in our Galaxy have been great hunting grounds for pulsars, with 360 discoveries in a total of 46 clusters\footnote{For an up-to-date list, see  \url{https://www3.mpifr-bonn.mpg.de/staff/pfreire/GCpsr.html}.}. 
The pulsar populations in each globular cluster are different in size, with denser and nearby GCs having generally more known pulsars; however, the characteristics of the pulsar populations of similarly dense GCs can also different as well, and sometimes strikingly so. Many of the denser GCs, especially core-collapsed clusters, have a large predominance of isolated pulsars, like NGC 6522 \citep{Abbate2023}, NGC 6624 \citep{Abbate_2022}, NGC 6752 \citep{Corongiu_2024}, M15 \citep{Wu_2024,Dai_2025}, NGC 6517 \citep{Lynch_2011,Yin_2024} and Terzan 1 \citep{Singleton_2024}. Other very dense non-core collapse clusters are either dominated (as in the case of NGC 1851, \citealt{Ridolfi_2022}, and Terzan 5, \citealt{Ransom2005,padmanabh_2024}) or entirely populated (as M62,\citealt{Vleeschower_2024})  by binary pulsars.

For the lower-density GCs, the pulsar population resembles the Galactic population of MSPs, which is dominated by binary systems with low-mass companions in low-eccentricity orbits. Examples of this are GCs like M3 \citep{Li_2024}, M5 \citep{Zhang_2023}, M13 \citep{Wang_2020}, M53 \citep{Lian_2023,Lian_2025b} and M71 \citep{Lian_2025a}. This is thought to be a consequence of the low stellar interaction rate per binary in these clusters \citep{Verbunt+Freire}: once a neutron star - main sequence star system is formed (in globular clusters, this happens via exchange encounters), it evolves undisturbed into a low-mass X-ray binary and then into a MSP system similar to those seen in the Galaxy.

Apart from M53 and M71 \citep{Lian_2023,Lian_2025a}, NGC~6749 is the least dense GC where pulsars have been discovered.
Some of its parameters are provided in Table~\ref{table:timing_params}. It has one known pulsar, PSR~J1905+0154A
(henceforth \psr), which was discovered in a survey of globular clusters made with the Arecibo 305-m radio telescope \citep{Hessels_2007} using the L-wide receiver and the Wideband Arecibo Pulsar Processor (WAPP, \citealt{dowd_2000}) autocorrelator, which had a bandwidth of 100 MHz. This was centred at 1175 and 1475 MHz, depending on the expected DM for the globular cluster.

The pulsar was found in an observation made on 2002 August 8.
It has a spin period of 3.199 ms and a DM of $\sim$193.7 \dmunit; it was shown by subsequent observations to be in a binary system with an orbital period of 0.81 days and a projected semi-major axis of 0.59 s. This implies a mass function of $3.3\, \times \, 10^{-4} \, \msun$, which, if we assume a pulsar mass of $1.35 \, \msun$ and a median orbital inclination of $60 \deg$, results in a median value of the companion mass of $0.103 \, \msun$. The orbit had no detectable eccentricity, furthermore no eclipses of any sort were detected, although it should be kept in mind that the orbital coverage near superior conjunction was relatively poor. For these reasons, the companion was thought to be a WD.

Despite many follow-up observations of NGC~6749 with the Arecibo telescope (described below), which resulted in detections of the pulsar in all observations, only a partial solution could be derived \citep{Hessels_2007}. This partial solution restricted the position to be within 1\arcmin of the
cluster centre, a strong indication of the association of the pulsar with this GC.
We have carried out later attempts at connecting the data set using {\Dracula}\footnote{See \url{https://github.com/pfreire163/Dracula}.} \citep{Freire_Ridolfi_2018}; but this resulted in many possible solutions, confirming that the combination of density and precision of the pulsar times of arrival (ToAs) obtained from the Arecibo timing was not enough to determine the rotation count unambiguously.

A second candidate, NGC~6749B, was detected in a single observation, made on 2003 October 9, with a spin period of 4.960 ms and a DM of 192 cm$^{-3}$ pc, but it was not confirmed in any other observation \citep{Hessels_2007}.

More recently, the Five Hundred meter Aperture Spherical Telescope (FAST) has resumed observations of NGC~6749. The increased sensitivity of FAST compared to Arecibo made the discovery of additional pulsars in this GC a likely event, however, as will be described here, no additional pulsars have been found to date. Importantly, NGC~6749B was not confirmed in these observations either.

Despite this, these FAST observations were useful because they allowed a detailed characterisation of \psr. In this paper, we present a joint analysis of the FAST and Arecibo data on this pulsar. The observations and their processing, which include a search for additional pulsars in the FAST observations, are described in section~\ref{sec:observations}.
The 20-year timing solution of \psr\, based on both FAST and Arecibo timing is presented in section~\ref{sec:timing}, in this section we also discuss some of the timing parameters. The multi-wavelength follow-up of \psr\, is presented in section~\ref{sec:optical}, the highlight is the optical identification of the pulsar's companion. We summarise our findings in section~\ref{sec:conclusions}.

\section{Observations and data processing}
\label{sec:observations}

\subsection{Early radio observations at Arecibo}

The Arecibo data used in this work is the same the data used and listed by \cite{Hessels_2007} to characterise \psr. These result from 21 observations taken with the L-wide between 2004 March 06 and 2007 April 20. During this campaign, four WAPP back-ends had become available, three of which were used. Within the band of the L-wide receiver (1120-1730 MHz), these were normally centred at frequencies of 1170, 1420 and 1520 MHz; the frequency gap between the 1170 and 1420 MHz bands was to avoid persistent radio frequency interference (RFI) between 1220 and 1360 MHz, no data was taken using the fourth WAPP at 1620 MHz because of persistent RFI above 1570 MHz. The lag data from each autocorrelator were then Fourier transformed into ``Filterbank'' data, with 512-channel total intensity spectra recorded every 128 $\mu$s. At the DM  of \psr\, ($195 \, \rm pc \, cm^{-3}$), the intra-channel dispersive smearing is $110\,\upmu\rm s$, adding the time resolution in quadrature we obtain an effective time resolution of $169 \, \upmu\rm s$.

\cite{Hessels_2007} describe how these data were then folded and how the resulted pulse profiles were used to estimate pulse times of arrival. Posteriorly, and in a similar way to the analysis done for the pulsars in M5 \citep{Zhang_2023}, we shifted the frequency channels by 1/2 of the channel width (100 MHz / 512 / 2 = 0.098 MHz). This depends on the band is being down-converted: for the first set of observations (MJD 53070 - 53247), we subtracted half a channel bandwidth from the TOA frequency, for all subsequent Arecibo data there was a band inversion, which them leads to the addition of half a channel to the reported TOA frequencies. This significantly reduced the residual rms for these data.

\subsection{Recent observations with FAST}

NGC 6749 was observed by FAST nineteen times, the first observation happening on
2019 September 14 and the last one on 2024 November 03. These observations are listed in
Table~\ref{table:FAST_observations}. The total observing time was
about 18 h.

\begin{table}

\caption{List of observations of the globular clusters NGC 6749 with FAST.}
\centering 
\begin{tabular}{rr}
\hline
Date & Observation length (s)\\
\hline
2019-09-14 &    1800 \\
2019-10-26 &	3600 \\
2019-12-10 &   10800 \\
2020-08-31 &	1800 \\ 
2020-09-21 &	7200 \\
2021-01-13 &     900 \\
2021-10-23 &    3000 \\
2022-11-19 &    3000 \\
2022-12-19 &    3000 \\
2023-01-22 &    7200 \\
2023-01-30 &    7200 \\
2023-03-02 &    3600 \\
2023-04-02 &	3600 \\
2023-05-06 &    4200 \\
2023-10-05 &    4800 \\
2023-11-03 &    3000 \\
2024-03-10 &    6600 \\
2024-05-04 &    8400 \\
2024-11-03 &	4200 \\
\hline
\label{table:FAST_observations}
\end{tabular}
\tablefoot{
The observation on 2023 March 02 produced no detection. The last observation was taken with 
full Stokes data.}
\end{table}

All of these observations share the same basic setup parameters: they used the central beam of the FAST 19-beam receiver, which has a bandwidth
of 500 MHz centred at a frequency of 1250 MHz. This band was divided in a total of 4096 channels, hence each channel had a bandwidth of $\sim$122.07 kHz. The 4096-channel power spectra were accumulated and written to disk every 49.152~$\upmu$s. At the DM of \psr\, the channel bandwidth results in an intra-channel dispersive smearing of 101~$\upmu$s at 1250 MHz. Adding the time resolution in quadrature, we obtain an effective time resolution of 112~$\upmu$s. In all these observations we obtained only total intensity data (Stokes I), with the exception of the last one, where full Stokes data were taken.
The 2021-10-23 observation was conducted in Snapshot mode, where the observation was split into four equal segments, and only the third segment was pointed at the cluster centre.

\subsection{Search for additional pulsars in NGC 6749 in the FAST data}
\label{sec:search}

All the FAST observations were thoroughly searched for new pulsars using the standard acceleration-search procedures adopted in the FAST globular cluster survey \citep[see, e.g.,][]{Lian_2025a}.
To improve sensitivity to pulsars in compact binary systems, we also performed segmented searches following the general strategy adopted in recent FAST globular cluster searches for compact binaries \citep{Yin_2025}.
To account for the varying observation durations across epochs, all data were split into equal segments of approximately 900--3600 s for acceleration searches, using the \textsc{MOSS} script \footnote{Multiple Observation Segment Search (MOSS) for Pulsars: \url{https://github.com/ydejiang/MOSS}} \citep{Yin_2025}.
Each segment was then searched independently with an acceleration of $z_{\rm max}=200$.  
Apart from the re-detection of \psr, these searches yielded neither any additional pulsar detections nor a confirmation of NGC~6749B.

To further enhance the sensitivity to very faint pulsars, we also divided all observations into 3600 s segments and incoherently stacked their power spectra. This method resulted in a very high-S/N detection of \psr, but again revealed no evidence for any additional pulsars in the cluster.

\subsection{Optical observations with the Hubble Space Telescope}

NGC~6749 was observed with HST for the first time in 2024 using the Wide Field Camera of the Advanced Camera for Surveys, as part of the Hubble Missing Globular Cluster Survey (GO: 17435, PI: Massari, see \citealt{Massari_2025}). Observations consist of F606W and F814W exposures, with eight images per filter and exposure times ranging from 80 s to 699 s. PSF photometry was performed on the calibrated flc images following the well-establish workflow outlined in various papers focused on high-precision astrometry and photometry with HST images \citep[e.g.,][]{2017BelliniwCenI,libralato22}. The procedure has been optimised for the detection of faint and blue sources, such as typical MSP companions \citep[for full details, see][]{Rosignoli_2025}. Photometry was calibrated to the VEGAMAG system using appropriate zero-points and aperture corrections. Following the prescriptions of \citet{Kozhurina15}, source positions were aligned to the International Celestial Reference System (ICRS) by cross-matching with Gaia DR3 \citep{GaiaCollaboration2023}, achieving a 1$\sigma$ astrometric accuracy with a root-mean-square residual of $\sim15$ mas. Since the cluster lies in a region of the Galaxy strongly affected by differential extinction \citep[see, e.g.,][]{cadelano_ngc6256a,cadelano24}, magnitudes were corrected for this effect using the widely adopted technique presented by \cite{milone12}.

\section{Timing solution of \psr}
\label{sec:timing}

\begin{figure}
\centering
	\includegraphics[width=\columnwidth]{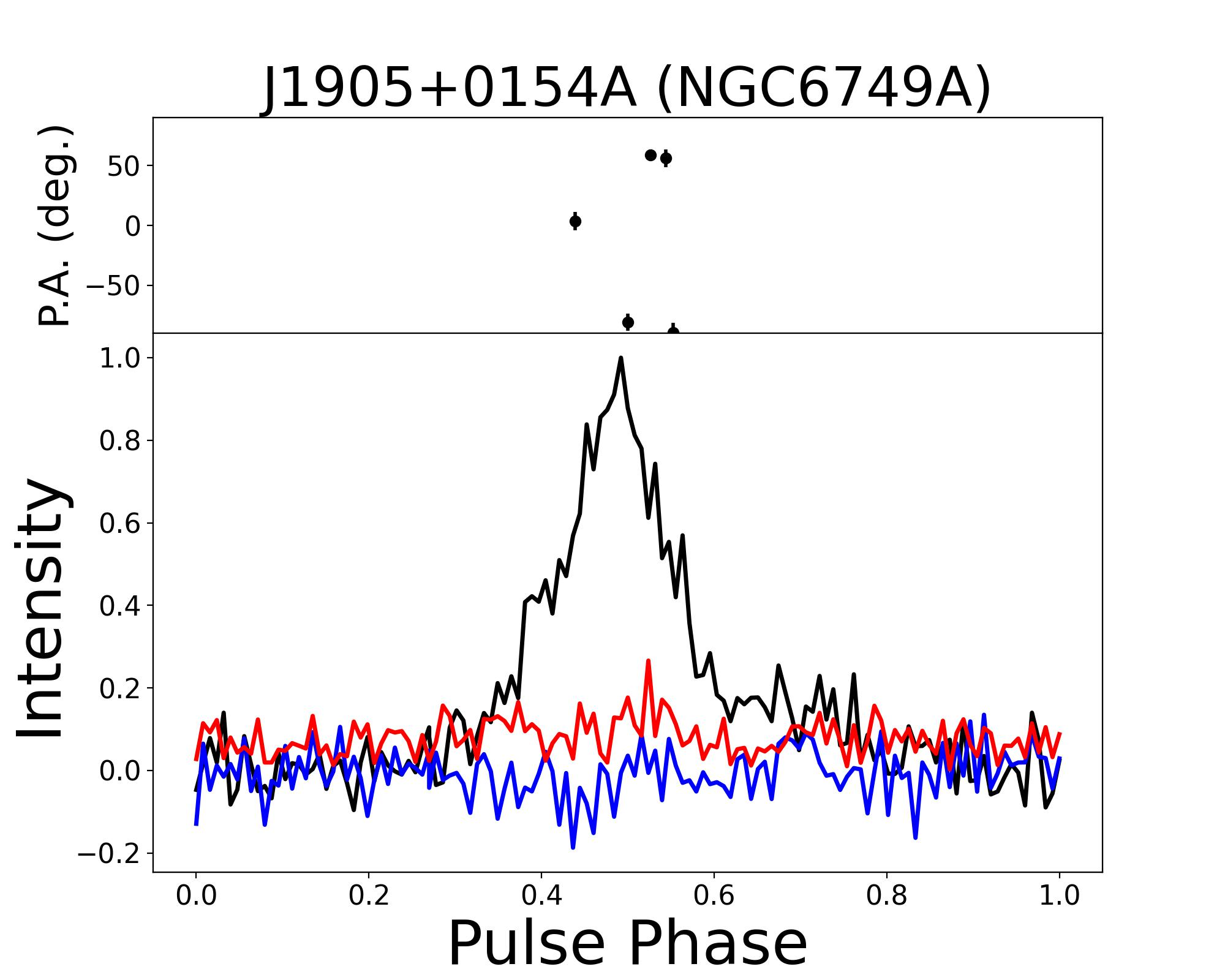}
	\caption{Polarized profile for NGC~6749A at L-band obtained from polarimetric data.}
        \label{fig:profile}
\end{figure}

As in the previous Arecibo observations, \psr\, can be detected reliably in our FAST observations. A polarimetric profile, resulting from a single polarimetric observation taken on 2024 November 03, is displayed in Fig.~\ref{fig:profile}. This shows the broad profile of the pulsar at L-band, this and the low S/N of the detection result in the relatively low precision of the timing. No significant polarisation, either liner or circular, were detected in the signal.

From the ToAs of \psr, a unique timing solution has been derived using the FAST data, the first time for this pulsar. This was then
verified to unambiguously connect the early Arecibo ToAs, thus implying we now have an ephemeris with a baseline of 20 years. The timing solution derived taking into account all available data is presented in Table~\ref{table:timing_params}. In this analysis, we used the \tempo\, timing package\footnote{\url{https://tempo.sourceforge.net}}.
This solution has a reduced $\chi^2$ of 1.06, which was derived as follows: for a preliminary version of the final timing solution, all parameters were kept fixed, and the reduced $\chi^2$ of each data set were estimated using \tempo, then the ToA uncertainties were increased by a factor $h$ in order to obtain a reduced $\chi^2$ of 1.0 for each data set. For each data set, the respective $h$ factor was then taken into account in the derivation of the full solution, by refitting all relevant timing parameters. 

\begin{table}
\caption{Parameters of the globular cluster NGC 6749 and the \psr\, system.
}
\centering 
\begin{scriptsize}
\begin{tabular}{lc}
\hline
\hline
\multicolumn{2}{l}{Globular cluster parameters}\\
\hline
Right ascension, $\alpha_{\rm GC}$ (J2000)\dotfill & 19:05:15.36 \\
Declination, $\delta_{\rm GC}$ (J2000)\dotfill & 01:54:3.6 \\
Distance, $D$ (kpc) \dotfill & $7.24^{+69}_{-57}$ \\
Proper motion in $\alpha$, \muraGC (\masy)\dotfill & $-2.829 \pm 0.028$  \\
Proper motion in $\delta$, \mudecGC (\masy)\dotfill  & $-6.006 \pm 0.027$ \\
Core radius (arcmin) \dotfill & 0.62 \\
Central velocity dispersion, $\sigma_V$ (\kms) \dotfill & 2.64 \\
Escape velocity, $V_{\rm esc}$ (\kms) \dotfill & 11.8 \\
$a_{\rm GC}$ ($10^{-9} \rm m \, s^{-2}$) \dotfill & $-0.218$\\
\hline
\multicolumn{2}{l}{Definitions of the timing solution}\\
\hline
Terrestrial Time Standard   \dotfill &  TT(BIPM2023)  \\
Time Units \dotfill    & TDB  \\
Solar System Ephemeris  \dotfill &  DE440  \\
Span of timing data (MJD) \dotfill & 53070 -- 60617 \\
Reference Epoch (MJD)\dotfill & 60378 \\
Number of TOAs \dotfill & 253 \\
Weighted rms residual (${\upmu} \rm s$) \dotfill & 27.8 \\
$\chi^2$ \dotfill & 235.3 \\
reduced $\chi^2$ ($\chi^2/n_{\rm free}$) \dotfill & 1.006 \\
Reduced 
Weighted rms residual (${\upmu} \rm s$) \dotfill & 27.8 \\
\hline
\multicolumn{2}{l}{Spin and astrometric parameters}\\
\hline
Right ascension, $\alpha$ (J2000)\dotfill & 19:05:14.69052(19) \\
Declination, $\delta$ (J2000)\dotfill & 01:54:48.274(8) \\
Proper motion in $\alpha$, \mura (\masy)\dotfill & $-2.52(24)$  \\
Proper motion in $\delta$, \mudec (\masy)\dotfill  & $-9.4(8)$ \\
Spin frequency, $\nu$ (Hz)\dotfill & 313.190896837(10) \\
First spin frequency derivative, $\dot{\nu}$ ($10^{-15}$\,Hz\,s$^{-1}$)\dotfill & $-$1.885(17) \\
Dispersion measure, DM (\dmunit) \dotfill  & 195.0(3.3) \\
First DM derivative, DM001 (\dmunit\,yr$^{-1}$) \dotfill  & 0.06(36) \\
Second DM derivative, DM002 (\dmunit\,yr$^{-2}$) \dotfill  & 0.012(21) \\
Third DM derivative, DM003 (\dmunit\,yr$^{-3}$) \dotfill  & 0.0054(9) \\
Fourth DM derivative, DM004 (\dmunit\,yr$^{-4}$) \dotfill  & 0.00069(13) \\
\hline
\multicolumn{2}{l}{Orbital parameters}\\
\hline
Orbital period, $P_\mathrm{b}$ (day)\dotfill & 0.8125524389(21) \\
Projected semi-major axis of pulsar orbit, $x \equiv a_\mathrm{p}\sin i/c$ (s) \dotfill & 0.5886568(41) \\
Time of ascending node, $T_{\rm asc}$ \dotfill & 60378.0292965(21) \\
$\epsilon_1 \equiv e \sin \omega$ \dotfill & $-$0.000031(13) \\
$\epsilon_2 \equiv e \cos \omega$ \dotfill & 0.000002(12) \\
Observed orbital period derivative, \pbdot (10$^{-12} \mathrm{s}\,\mathrm{s}^{-1})$ \dotfill & $0.4(6)$ \\
\hline
\multicolumn{2}{l}{Optical parameters of companion WD}\\
\hline
Reddening-corrected F606W magnitude, $m_{F606W}$ \dotfill & 26.08(6) \\
Reddening-corrected F814W magnitude, $m_{F814W}$ \dotfill & 24.78(5) \\
White dwarf temperature \dotfill & 11,600 - 14.800 \\
White dwarf mass ($\msun$) \dotfill & 0.17 - 0.19 \\
Cooling age (Gyr) \dotfill & 0.4 - 0.7 \\
\hline
\multicolumn{2}{l}{Derived parameters}\\
\hline
Galactic longitude of the cluster centre, $\ell$ (deg)  \dotfill & 36.201 \\
Galactic latitude of the cluster centre, $b$ (deg) \dotfill & $-2.205$ \\
DM Distance, $d_\mathrm{NE2001}$ (kpc) \dotfill & 3.4 \\
DM Distance, $d_\mathrm{YMW16}$ (kpc) \dotfill & 5.5 \\
DM Distance, $d_\mathrm{NE2025}$ (kpc) \dotfill & 3.8 \\
Offset of pulsar from the centre of the cluster (\arcmin) \dotfill & 0.76 \\
Offset of pulsar from the centre of the cluster (core radii) \dotfill & 1.25 \\
Projected distance of from the centre of the cluster (pc) \dotfill & 1.6 \\
Total proper motion, $\mu_{\text{T}}$ (\masy) \dotfill  & 9.7(7) \\
Position angle of proper motion, J2000, $\Theta_{\mu}$ ($\deg$) \dotfill & 195.0(18) \\
Proper motion relative to the cluster (\masy) \dotfill  &  3.5(8) \\
Transverse velocity relative to the cluster (\kms) \dotfill  & 123(34) \\
Spin period, $P$ (ms) \dotfill & 3.19294082330(10) \\
Spin period derivative, $\dot{P}_\mathrm{obs}$ ($10^{-20} \mathrm{s}\,\mathrm{s}^{-1}$) \dotfill & 1.922(17)  \\
Cluster contribution to $\dot{P}_\mathrm{obs}$ ($10^{-20} \mathrm{s}\,\mathrm{s}^{-1}$) \dotfill & $\pm$0.18  \\
Galactic acceleration contribution to $\dot{P}_\mathrm{obs}$ ($10^{-20} \mathrm{s}\,\mathrm{s}^{-1}$) \dotfill & $-$0.23\\
Proper motion contribution to $\dot{P}_\mathrm{obs}$ ($10^{-20} \mathrm{s}\,\mathrm{s}^{-1}$) \dotfill & $+$0.52 \\
Intrinsic spin period derivative, $\dot{P}_\mathrm{int}$ ($10^{-20}\, \mathrm{s}\,\mathrm{s}^{-1}$) \dotfill &  1.64$\pm$0.18 \\
Characteristic age (Gyr) \dotfill  & 2.8 - 3.5 \\
Surface magnetic field ($10^8$ G) \dotfill & 2.2 - 2.4\\
Orbital eccentricity, $e$ \dotfill & 0.000032(12) \\
Epoch of periastron, $T_{0}$ (MJD) \dotfill & 60378.667(55) \\
Longitude of periastron, $\omega \, (\deg)$ \dotfill & 282(25) \\
Mass function $f\, (\msun)$ \dotfill & 0.000331716(7) \\
Minimum companion mass, $M_\mathrm{c, min}$ \dotfill & 0.088 \\
Median companion mass,  $M_\mathrm{c, med}$ \dotfill & 0.103 \\
Orbital inclination for $m_c = 0.18 \, \msun$ \dotfill & 30.7 \\
\hline
\label{table:timing_params}
\end{tabular}
\tablefoot{
The timing parameters were derived using \tempo\ with the ELL1H binary model. The values in the parenthesis indicate the 1-$\sigma$ uncertainties on the last digit. The minimum and median companion masses and orbital inclination are estimated
assuming a pulsar mass of $1.35 \, \msun$. The linear distances and velocities relative to the centre of the GC assume the distance estimated for the GC.
}
\end{scriptsize}
\end{table}

\begin{figure*}
\sidecaption
     \includegraphics[width=12cm]{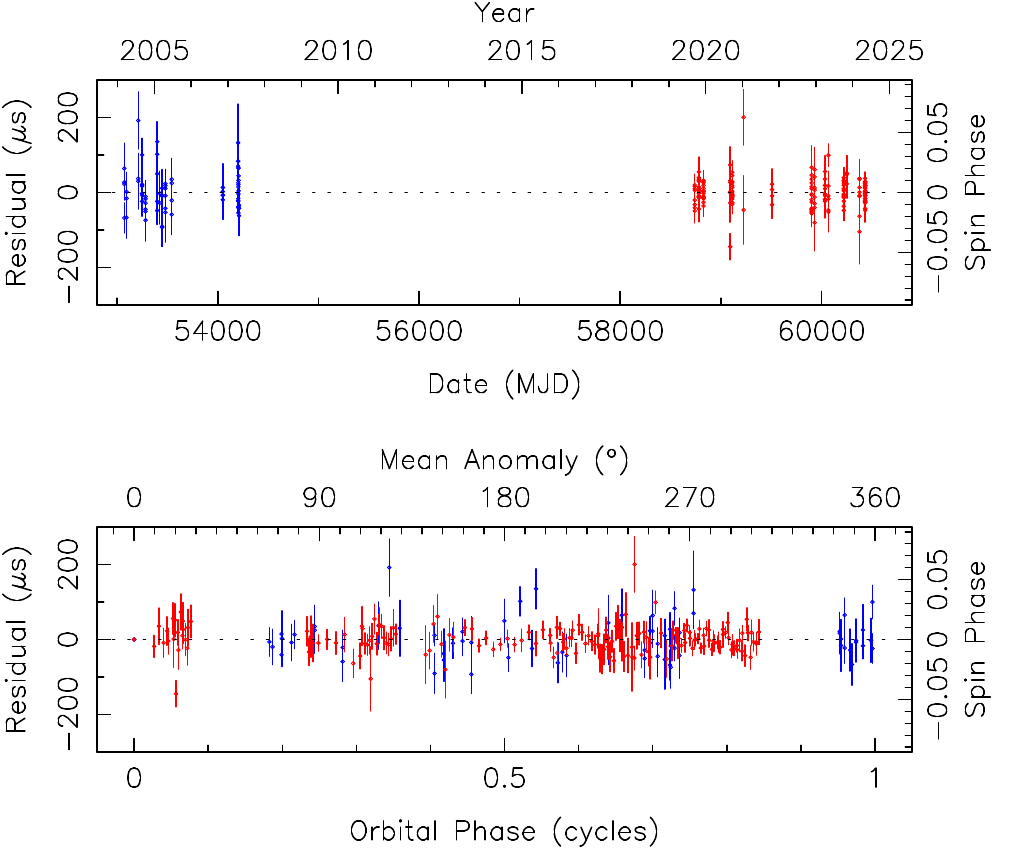}
     \caption{ToA residuals obtained with our set of ToAs and the timing solution in Table~\ref{table:timing_params}. The early ToAs (in blue) were obtained with the Arecibo observing setup; the latter (in red) with the FAST observing setup. The top plot shows the residuals as a function of time, the lower plot shows them as a function of orbital phase relative to ascending node. No trends are visible in the residuals, which means that the timing solution in Table~\ref{table:timing_params} provides a good description of the data. Furthermore, no eclipses are seen at superior conjunction, which occurs at an orbital phase of 0.25.}
     \label{fig:residuals}
\end{figure*}

The timing residuals (ToA minus model prediction) are presented in Figure~\ref{fig:residuals}. No trends are visible in the residuals, either as a a function of time or orbital phase; this means that the timing solution in Table~\ref{table:timing_params} provides a good description of the data. The orbital model used was the ELL1H model \citep{Lange_2001,Freire_Wex}, which uses the Laplace-Lagrange parameters ($T_{\rm asc}, \epsilon_1, \epsilon_2$) to describe the orbit and the orthometric parameters of the Shapiro delay ($h_3$ and $\varsigma$) to describe the relativistic light propagation delay \citep{Shapiro_1964} in the spacetime of the binary pulsar.
Relative to the Damour-Deruelle timing formula \citep{DD_1986} and its parameters, the previous parameters avoid the strong correlations between the time of passage of periastron ($T_0$) and the longitude of periastron ($\omega$) observed for low-eccentricity systems and between the range ($r$) and shape ($s$) of the Shapiro delay observed for low-inclination systems. 

One of the features of this globular cluster (and its pulsar) is the large DM, a consequence of the large electron column density. As seen for other pulsars, the larger the DM, the larger are its variations. In this work, we model them as a Taylor expansion, with four DM derivatives. Using these reduced the value of $\chi^2$ from, successively, 284.9, 273.2, 270.0, 264.8 and 235.3. Adding further DM derivatives (up to the 9th) does not significantly decrease the value of $\chi^2$.

In what follows we discuss the significance of some of these timing parameters.

\subsection{Position and proper motion relative to the cluster's}

The position of the centre of NGC 6749, its distance and its proper motion listed in Table~\ref{table:timing_params} were  taken from \cite{Vasiliev_2021}. The core radius, central velocity dispersion and escape velocity were taken from \cite{Baumgardt_2018}. Using these values, we find that the pulsar is 0.76 arcminutes from the centre of the
cluster, or about 1.25 core radii; confirming the preliminary conclusion by \cite{Hessels_2007} on the pulsar being close to the GC centre. At a distance of 7.24 kpc, this corresponds to a projected distance of 1.6 pc.
This kind of distance is very typical of what has been found for other GCs where the pulsars are found
to have a dynamically relaxed distribution, i.e., where mass segregation has run its course (see e.g., \citealt{Freire_2017,Abbate_2018,Prager_2017,Zhang_2023,Li_2024,Vleeschower_2024,Lian_2023,Lian_2025a,Lian_2025b}). The precise position determined from timing allowed the identification of the companion in optical images of the GC, these will be discussed in detail in the next section.

However, the proper motion difference (pulsar $-$ globular cluster) in right ascension and declination are 0.31(24) \masy\ and $-3.4$(8) \masy\ respectively, for a total difference of 3.4(8) \masy.
This translates, at the distance to the GC, to a velocity difference of 116 $\pm$ 26 \kms, which has a significance of about 4.5 $\sigma$ and is certainly larger than the escape velocity from the centre of the GC, which is about 11.8 \kms. If confirmed, then the pulsar is not associated with the cluster; the probability of a chance coincidence with the GC is discussed in detail below.
However, one should keep in mind that DM variations introduce systematic biases in parameters that require long timing baselines like the proper motion.

\subsection{Acceleration, spin-down and orbital parameters}

For pulsars in GC, the observed spin period derivative is given by:
\begin{equation}
\label{eq:pb-dot}
\left(  \frac{\dot{P}}{P}  \right)_{\rm obs} = \left(  \frac{\dot{P}}{P}  \right)_{\rm int} + \frac{a_{\rm Gal} + a_{\rm GC}}{c} + \frac{\mu^2 d}{c},
\end{equation}
where $\dot{P}_{\rm int}$ is the intrinsic spin period derivative, $a_{\rm Gal}$ and $a_{\rm GC}$ are the accelerations of the binary system in the gravitational fields of the Galaxy and the globular cluster respectively, and the last term is the Shklovskii effect \citep{Shklovskii_1970}. A similar equation can be written for \pbdot.
Generally the terms $a_{\rm Gal}/c$ and $\mu^2 d/c$ are of the same order, and in this case partially cancel each other; they are listed in Table~\ref{table:timing_params}, where the Galactic acceleration term for the pulsar and GC is estimated using the \cite{McMillan2017} galactic acceleration model. 

The unknown contributions arise from $\dot{P}_{\rm int}/P$ and $a_{\rm GC}/c$. 
For $a_{\rm GC}$, we can calculate the extremes at the position of the pulsar using a simple analytical King model \citep{Freire_2005}, see Table~\ref{table:timing_params}.
In most GCs, $a_{\rm GC}$ is the dominant term, however, given the very low density of the core of NGC 6749 the dominant term is $\dot{P}_{\rm int}/P$, with $a_{\rm GC}/c$ contributing less than 10\% of the observed $\dot{P}_{\rm int}/P$, and unusually, by an amount that is similar to that of $a_{\rm Gal}/c$ and $\mu^2 d/c$ . This situation is very similar to what has been observed for other GCs like M53 \citep{Lian_2023} and M71 \citep{Lian_2025a}. This means that, after taking into account all other terms, and considering the extremes $a_{\rm GC}$, we can deduce $\dot{P}_{\rm int}$ with about 10\% accuracy; from these, we derive a characteristic age between 2.8 and 3.5 Gyr and a magnetic field between 2.2 and $2.4\, \times \, 10^8\, \rm G$. These numbers are very typical of the Galactic MSP population. They are nonetheless interesting since, in general, they cannot easily be measured in more massive GCs, unless one has a very precise measurement of the acceleration of the binary via \pbdot\ \citep[see e.g.][]{Freire_2017,Dutta_2025} or the GC acceleration is much smaller than $\dot{P}_{\rm int}$ \citep{Abbate_2022,Lian_2023,Zhou_2024,Wu_2024,Lian_2025a}.

Re-writing eq.~\ref{eq:pb-dot} for \pbdot, we obtain a prediction of $0.06(4) \times 10^{-12} \rm \, s\, s^{-1}$ for the contribution of the kinematic terms to $\dot{P}_{\rm b}$. According to GR, and assuming pulsar masses of the order of $1.4\, \msun$ and companion masses of the order of $0.18 \, \msun$ (see section~\ref{sec:optical}) the contribution from orbital decay due to the emission of gravitational waves for this system is an order of magnitude smaller than this prediction; this is taken into account in this estimate. The observed value of \pbdot\, is, within its large error bar, consistent with this prediction; however, its uncertainty is still an order of magnitude more than the predicted effect.

\section{Multi-wavelength follow-up}
\label{sec:optical}

\subsection{X-rays}
From the HEASARC archive\footnote{\url{http://heasarc.gsfc.nasa.gov/db-perl/W3Browse/w3browse.pl}}, we find the deepest X-ray limits at \psr's timing position from a 4.7 ks Swift/XRT observation on Oct. 30, 2009, which detected no source near the pulsar position. Assuming a power-law of photon index 2, $N_H=7\times10^{21}$ cm$^{-2}$, and $d=7.9$ kpc, we estimate a rough limit of $L_X$(0.5-10 keV)=$4\times10^{32}$ erg/s (C. Heinke, priv. comm.). The X-ray luminosities of MSPs with He WD companions are typically $10^{30-31}$ erg/s \citep{Zhao2022}, so the non-detection of \psr\, is not surprising.

\begin{figure}
	\includegraphics[width=\columnwidth]{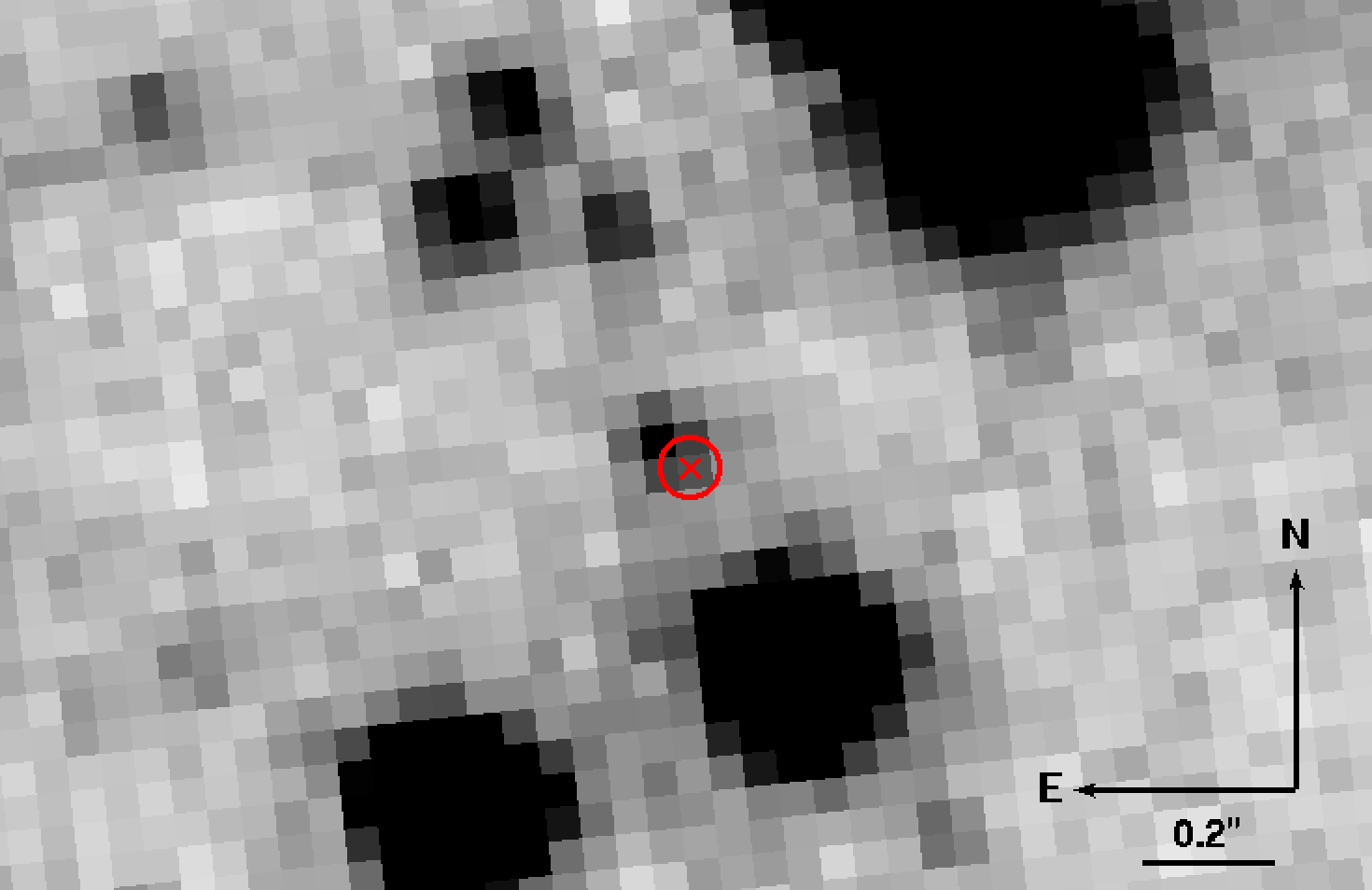}
	\caption{Finding chart of the region surrounding NGC~6749A in the F606W filter. The red cross marks the pulsar position, while the circle represents the 3$\sigma$ uncertainty radius from the combined optical and radio positional error. The only detected star within this region is the candidate companion to the pulsar.}
        \label{counterpart}
\end{figure}

\begin{figure*}
\centering
\sidecaption
	\includegraphics[width=12cm]{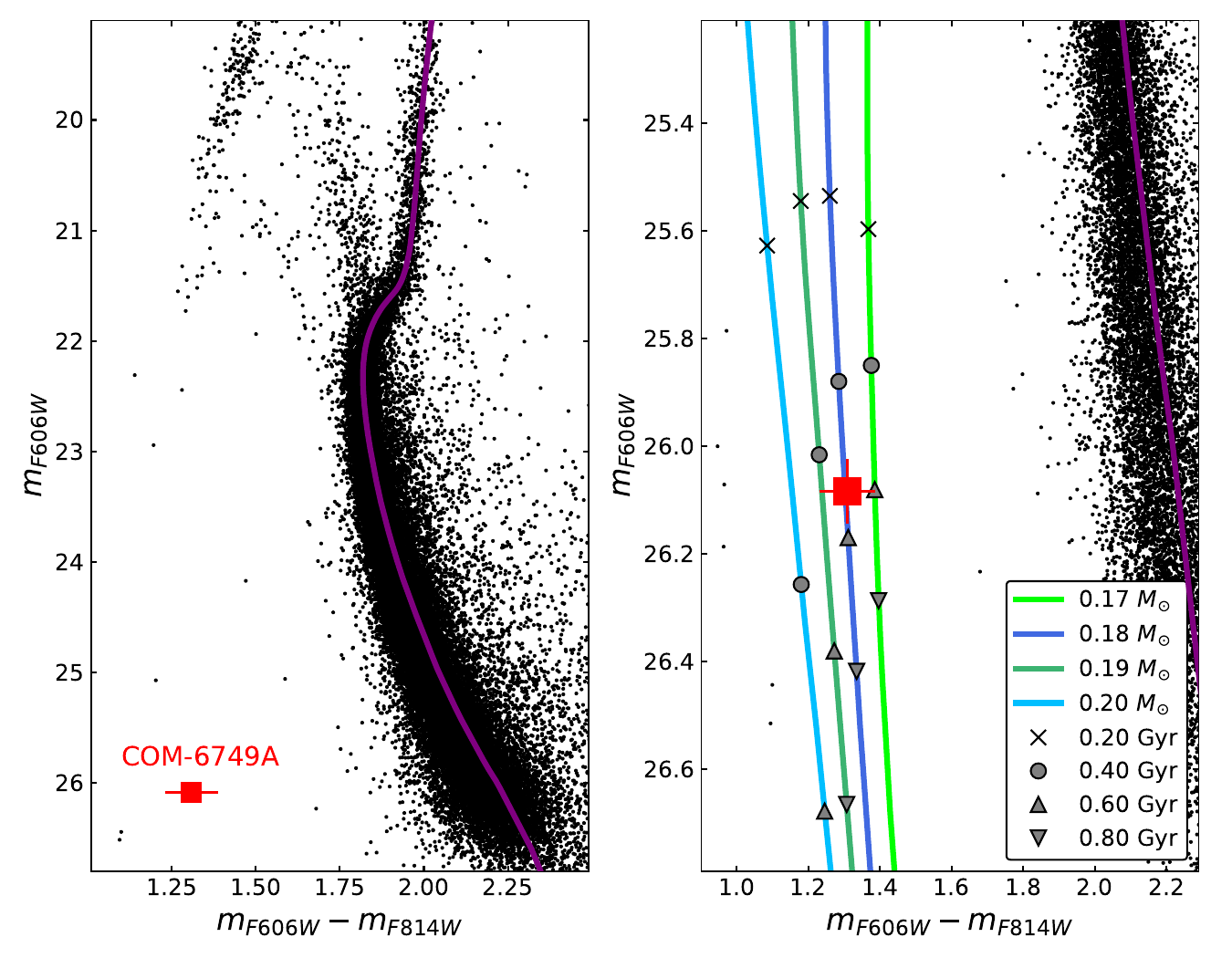}
	\caption{Left-hand panel: Colour-magnitude diagram of NGC6749 in a combination of the F606W and F814W. The red square is the position of the counterpart to NGC6749A. The purple curve is a 13 Gyr isochrone calculated at the cluster metallicity, distance and extinction. Right-hand panel: same as the left-hand panel, but zoomed in the companion CMD position. The coloured curves are He WD cooling tracks from \citet{Istrate2014,Istrate2016}. The tracks are for WD with masses in the range between $0.17\msun$ and $0.2\msun$, with decreasing masses from left to right, as reported in the legend. Different points are also highlighted with different markers along the tracks corresponding to different cooling ages.}
        \label{companion}
\end{figure*}

\subsection{Optical identification of the companion}

To identify the optical counterpart of the binary pulsar, we inspected all the stellar sources within a $1\arcsec \times 1\arcsec$ region surrounding the pulsar position. As shown in Figure~\ref{counterpart}, there is a star compatible with the pulsar displaced by only 0.04\arcsec from the pulsar position, thus within $3\sigma$ the combined optical and radio uncertainties. In the colour-magnitude diagram (CMD), the star is located in a blue region, with a colour similar to that of blue horizontal branch stars, consistent with the expected location of WDs. In fact, this is the only clearly detected WD in the cluster, which is not surprising given its distance and the significant extinction, which makes the detection of hot stars particularly challenging. The good positional coincidence and the CMD location, indicative of a WD nature, strongly suggest that this source is indeed the companion to NGC~6749A. The differential reddening-corrected magnitudes of the companion are: $m_{F606W} = 26.08 \pm 0.06$ and $m_{F814W} = 24.78 \pm 0.05$. 

The multi-band magnitudes can be used to infer the physical properties of the WD, such as its mass, surface temperature, and cooling age, by comparison with appropriate He WD cooling tracks. We adopted the models from \citet{Istrate2014,Istrate2016}, which follow the evolution of an NS binary through the entire mass transfer phase, the proto-WD stage, and the WD cooling phase. These tracks have been extensively used in our previous studies to characterise MSP companions \citep[e.g.,][]{Cadelano2020_m13,chen23,Etorre_2025}. Figure~\ref{companion} shows these cooling tracks along with a 13 Gyr isochrone (purple solid curve) extracted from the BaSTI database \citep{hidalgo2018,pietrinferni21}, assuming a cluster metallicity of $[Fe/H] = -1.6$ \citep{harris1996,harris2010}. We adopted a cluster distance modulus of $(m-M)_0 = 14.40$ and an average colour excess of $E(B-V) = 1.37$, in reasonable agreement with the values reported by \citet{harris1996,harris2010}. As shown in the figure, the models successfully reproduces the entire cluster evolutionary sequence, from the red-giant branch down to the faint main sequence, also confirming the He WD nature of the companion star. However, the companion is a faint source detected in only two optical filters in a challenging environment characterised by strong differential reddening. Given these limitations, we conservatively infer its properties by bracketing its CMD position within the range of expected magnitudes and colour from the cooling tracks. As shown in the right panel of Figure~\ref{companion}, its location in the CMD is consistent with a He WD with a mass of $0.17\msun - 0.19\msun$, a cooling age of $0.4 - 0.7$ Gyr, and a surface temperature of $11.600 - 14.800$ K. A more detailed analysis, similar to that presented in \citet{Cadelano2019,Cadelano2020_m13}, would require additional multi-band observations, particularly in the blue and near-UV bands.

This companion mass and the mass function of the system imply a low orbital inclination: assuming a companion mass of $0.18\, \msun$ and pulsar masses of 1.35 and $2.0 \, \rm \msun$, the orbital inclinations are 30.7 and 40.2 degrees.

\section{Summary and conclusions}
\label{sec:conclusions}

In this work, we have presented the results of recent FAST observations of NGC~6749. 
Apart from the previously known pulsar, \psr, we have not been able to detect any additional pulsars in this GC. In particular, we have not been able to confirm NGC~6749B. The reasons for this are not clear. It is possible it is not a real pulsar, but it is also possible that it is an eclipsing binary that appears very rarely. Alternatively, it might be located outside the narrow FAST beam, but still be located at the outer margin of the wider Arecibo beam. 

The previously known pulsar, \psr, was detected consistently, from this we could obtain a timing solution for this binary pulsar. This timing solution allowed the connection of the early Arecibo data, resulting in a timing baseline of 20 years.

The acceleration of the system in the field of the GC is so small that even with 20 years of data, its effect on the orbital period derivative remains undetected; nevertheless the precise measurement of the spin period derivative and the small expected accelerations result in a relatively well estimated value for the spin-down of the pulsar, one of the few well-measured cases in GCs; this is small and results in a characteristic age between 2.8 and 3.5 Gyr and a magnetic field at the surface between 2.2 and $2.4 \times 10^8\,$G, implying a relatively normal MSP. Nevertheless, this is one of the few intrinsic spin-down ages and characteristic ages measured for pulsars in globular clusters.

The precise position of the pulsar allowed the identification of one of the few candidate WDs in this cluster as the pulsar's companion; thus confirming it as a WD. There are three objects near the CMD position of the companion ($25 < m_{606} < 26.5$ and colour < 1.25) which might be WDs or not; they could also be sources with bad photometry or sources not associated with the cluster, such as quasars. Thus, the companion of \psr\, is the only confirmed WD in this GC. The latter's photometric characteristics are consistent with a He WD with a mass of $0.17\msun - 0.19\msun$, a surface temperature of $11.600 - 14.800$ K and a cooling age of $0.4 - 0.7$ Gyr. According to Eq. 1 in \citep{Istrate2014}, for the mass of this WD companion, the proto-WD age, i.e., the phase after Roche-lobe detachment during which the WD contracts before entering the cooling sequence, lasted 0.6–1.2 Gyr, implying a time between 1.0 and 1.9 Gyr since Roche lobe detachment. This age is consistent with (i.e., of a similar magnitude, but not larger than) the characteristic age of the pulsar. Using the equation for the change of the spin period with time,
\begin{equation}
p(t) = P \left[ (n-1) \frac{\dot{P}}{P} t + 1  \right]^{\frac{1}{n -1}},
\end{equation}
where $n$ is the braking index. A constant magnetic field ($n = 3$) and the nominal value of $\dot{P}_\mathrm{int}$ would imply that, at the time of Roche lobe detachment, the spin period was between $1.98$ and $2.62\, \rm ms$. Such values are well within the range of observed MSP spin periods.
An interesting aspect of this measurement is that it can be used to constrain the braking index.
In particular, for $n = 4$ and $t = -1.9\, \rm Gyr$, the initial spin period would be 1.35 ms, which would be shorter than any observed to date \citep{Hessels_2006}.

The only unanswered question raised by the current work is the large difference between the proper motion of the pulsar and the globular cluster, at least in declination. If confirmed, this could imply that the system is not associated with the cluster. Given the possibility of systematics due to DM variations, we consider this unlikely, especially given the close proximity to the centre of the GC, 0\farcm76. However, it cannot be excluded, for two additional reasons: first, for the coordinates of this pulsar and its DM, the three most used DM models (NE2001, \citealt{ne2001}, YMW16, \citealt{YMW2016} and NE2025, \citealt{NE2025}) 
predict the distances that range from 3.4 to 5.5 kpc (see Table~\ref{table:timing_params}); these fall systematically short of the estimated distance to the globular cluster of $7.24^{+69}_{-57}\, \rm kpc$. However, given the large uncertainties of these models, which also fail to predict the distances to other globular clusters from the DMs of well-established pulsar members of those clusters, this is not a a strong argument against the association. Second, for a region of the sky of about 60 square degrees around the position of \psr\, with $|b| < 3^\circ$ and $31^\circ < \ell < 41^\circ$, the ATNF catalogue \citep{Manchester_2005} currently lists 359 pulsars, with 49 of them having spin periods smaller than 10 ms.
This represents pulsar densities of about 6 and 0.8 per square degree respectively.
Thus, the possibility of having a pulsar, and in particular a millisecond pulsar, within any particular beam of the Arecibo telescope (which at L-band had a full width at half maximum of about 3 arcminutes, or 0.05 degrees) is about $1.2\%$ and $0.16\%$ respectively.
These probabilities increase if we consider the great depth of the Arecibo observation that found this pulsar, which had an integration time of several hours.

If the pulsar were located at these smaller distances, the WD companion would be significantly fainter, which would imply larger WD masses (0.20 - 0.25 $\msun$) and possibly much smaller cooling ages. These would still be consistent with the observed mass function, but would result in lower orbital inclinations: assuming a pulsar mass of $1.35 \, \msun$, companion masses of 0.20 and $0.25\, \msun$, the inclinations would be 27.6 and 22.3 degrees.

Future measurements should significantly improve the precision and the accuracy of the proper motion measurement and clarify the association.

\begin{acknowledgements}
PCCF gratefully acknowledges continuing support from the Max Planck Society and productive discussions with Michael Kramer.
Y. D. acknowledges support and guidance from his PhD supervisor, Professor Xingjiang Zhu.
Z. Z. acknowledges support from the Science Basic Research Program of Shaanxi (Program No. 2024JC-YBQN-0036).
We thank Tong Liu for providing the polarized pulse profile of NGC 6749A.
This work made use of the data from FAST (Five-hundred-meter Aperture Spherical radio Telescope). FAST is a Chinese national mega-science facility, operated by National Astronomical Observatories, Chinese Academy of Sciences. While taking data for this work, the Arecibo Observatory was operated by SRI International under a cooperative agreement with the U.S. National Science Foundation (NSF; AST-1100968), and in alliance with Ana G. M\'endez-Universidad Metropolitana, and the Universities Space Research Association.
This work has made use of data from the Hubble Treasury programme entitled "The Hubble - Missing Globular Cluster Survey" (GO-17435, PI: D. Massari). Data products are available at \url{https://www.oas.inaf.it/en/research/m2-en/mgcs-en/}.
 \end{acknowledgements}

\end{document}